# Nonlocal Position Changes of a Photon Revealed by Quantum Routers

Avshalom C. Elitzur[1,2], Eliahu Cohen[3,1,*], Ryo Okamoto[4,5], Shigeki Takeuchi[4]

[1]Iyar, The Israeli Institute for Advanced Research, POB 651, Zichron Ya'akov 3095303, Israel
[2]Institute for Quantum Studies, Chapman University, Orange, CA 92866, USA
[3]Physics Department, Centre for Research in Photonics, University of Ottawa, Advanced Research Complex, 25 Templeton, Ottawa ON Canada, K1N 6N5
[4]Kyoto University, Department of Electronic Science and Engineering, Kyoto Daigaku-Katsura, Nishikyo-ku, Kyoto 615-8510, Japan
[5]JST, PRESTO, Kawaguchi, Saitama 332-0012, Japan

*Correspondence to eli17c@gmail.com

*Since its publication, Aharonov and Vaidman's three-box paradox has undergone three major advances: i). A non-counterfactual scheme by the same authors in 2003 with strong rather than weak measurements for verifying the particle's subtle presence in two boxes. ii) A realization of the latter by Okamoto and Takeuchi in 2016. iii) A dynamic version by Aharonov et al. in 2017, with disappearance and reappearance of the particle. We now combine these advances together. Using photonic quantum routers the particle acts like a quantum "shutter." It is initially split between Boxes A, B and C, the latter located far away from the former two. The shutter particle's whereabouts can then be followed by a probe photon, split in both space and time and reflected by the shutter in its varying locations. Measuring the former is expected to reveal the following time-evolution: The shutter particle was, with certainty, in boxes A+C at $t_1$, then only in C at $t_2$, and finally in B+C at $t_3$. Another branch of the split probe photon can show that boxes A+B were empty at $t_2$. A Bell-like theorem applied to this experiment challenges any alternative interpretation that avoids disappearance-reappearance in favor of local hidden variables.*

## Introduction

How literally should quantum formalism be taken? Ever since its advent, it has predicted phenomena which, by classical physics, are inconceivable. In response, several interpretations proposed that this is not the case, but it is rather some local hidden variables that operate underneath. Other advances, however, in the form of new mathematical proofs and experiments, proved that the bare formalism, no matter how odd, provides an accurate description.

Such is the familiar state revealed by the double-slit experiment, featuring in every introductory text on QM, to which Feynman [1] has referred as presenting the theory's core mystery. Consider its refined version. When a single particle traverses a Mach-Zehnder Interferometer (MZI), its state is given by the simple expression

$$|\psi\rangle = \frac{1}{\sqrt{2}}(|1\rangle + |2\rangle), \tag{1}$$

which, in everyday language, reads: *The particle traverses in some sense both paths 1 and 2, assuming a definite path only under position measurement taken during its passage.*

Upon position measurement, this inevitably invokes the notorious measurement problem and the contentious "collapse." Understandably, then, attempts were made to show that this superposition reflects only our subjective ignorance, while the particle itself goes on one definite side. Eventually, however, single-particle interference experiments indicated that both paths are simultaneously traversed. So much so, that even when one path is measured and yields no detection, Interaction-Free Measurement (IFM) ensues, destroying the interference effect just like a positive click. Quantum superposition, then, is the case just as (1) indicates.

Such was also the case with the EPR state. Its familiar two-spins version obeys the relation

$$|\psi\rangle = \frac{1}{\sqrt{2}}(|\uparrow\downarrow\rangle - |\downarrow\uparrow\rangle), \tag{2}$$

which, in everyday language, reads: *Each particle's spin (like the particle's position in Eq. 1) is both ↓ and ↑ along every direction, yet maximally correlated with the other, becoming definite in a certain direction for both particles only under measurement performed on either one of them.*

This state entails the (in)famous issue of quantum nonlocality. And here too, attempts to restore locality were proposed, again to be ruled out eventually, this time by Bell's [2] theorem and the subsequent tests. Further attempts were still made to find loopholes in these advances, all countered so far.

These advances have come full circle when Bell's theorem was extended by Hardy [3,4] also for the above, older case of the double-slit/MZI with a single particle. One extension is Hardy's paradox [3], where a particle and its anti-particle interact but do not annihilate. Their interference patterns indicate that nonlocal correlations have existed, *all along*, within each single wave-function [5]. In another setting [4], a single photon is entangled with two atoms placed along the two MZI's paths. These atoms thereby become an EPR-Bell pair, proving, again, that their nonlocal correlation were inherited from the *earlier* nonlocal correlation between the two halves of the single particle's wave-function.

In summary, the two most basic characteristics of QM merit being taken literally: *i*) A single particle propagates like a wave, somehow traversing all possible paths with its expanding front; and *ii*) Within a single wave-function, a local interaction with a measuring device instantly affects the entire wave-function, regardless of its spread in spacetime, while preserving relativistic causality.

The main contribution of this paper is a set of feasible experimental proposals for testing the nonlocality of a single particle, and specifically, its ability to disappear and reappear, with certainty, across various locations at certain instances. This unique dynamics is indicated by the particle's reflecting a probe photon from all these spacetime points, with the aid of photonic quantum routers. The extreme nonlocality in this setup is proved with a Bell-like setting.

The paper is organized as follows. In "Time to Accept Even Greater Oddities" we present a surprising prediction regarding the interaction of a probe photon with a pre- and post-selected shutter photon. In the next two sections we present the theoretical framework which rigorously derives this prediction from the standard quantum formalism. Then in "The Strong-Measurement Version of the Three Boxes Paradox and its Realization" we present an actual experiment, already performed, testing an earlier prediction of that framework, thereby setting the stage for the following. We proceed on to the newer predictions in "The Disappearing-Reappearing Particle" and in "Bell's Inequality Ruling Out the 'All Along' Alternative". Then come the main sections "The Proposed Realization of the Disappearing-Reappearing Particle Experiment" and "Experimental Setup and further Practical Considerations," where we present a feasible experimental setup for testing our predictions, with a few variants. We mention a stronger variant in "A Stricter Version: Measuring Both Appearances and Disappearances with the Same Probe Photon" and finally return to the question of nonlocality in "An EPR-Bell-Type Validation." We conclude with some general implications in "Discussion".

## Methods and Results

We submit that the same credibility assigned to the above states (1),(2) goes, for the same reasons, to even more intriguing phenomena recently derived from quantum theory. These can be summarized in equations (4),(6) below, already implicit in (1),(2) and presented in what follows with their own Bell-Hardy type proofs.

**Time to Accept Even Greater Oddities.**

The first intriguing prediction is Aharonov and Vaidman's [6,7], which has recently won empirical support by Okamoto and Takeuchi [8]. They considered the interaction (via a quantum router) between an incident (In) probe photon (pp) and another pre- and post-selected photon acting as a shutter (sp):

$$|i\rangle_{pp}|j\rangle_{sp} \to \begin{cases} |Ri\rangle_{pp}|j\rangle_{sp} & (i=j) \\ |Ti\rangle_{pp}|j\rangle_{sp} & (i \neq j) \end{cases}, \quad i = A, B, \quad j = A, B, C, \tag{3}$$

where $A, B, C$ denote three locations in space, *i.e.* three boxes, and $|Ri\rangle / |Ti\rangle$ stand for reflection/transmission from/through box $i$.

The probe photon was prepared in a superposition $|\text{In}\rangle_{pp} = \alpha_1 |A\rangle_{pp} + \alpha_2 |B\rangle_{pp}$, where $\alpha_1$ and $\alpha_2$ are two arbitrary complex coefficients satisfying $|\alpha_1|^2 + |\alpha_2|^2 = 1$, while the shutter photon was pre- and post-selected in $|\psi\rangle_{sp} = \frac{1}{\sqrt{3}}\left(|A\rangle_{sp} + |B\rangle_{sp} + |C\rangle_{sp}\right)$ and $|\phi\rangle_{sp} = \frac{1}{\sqrt{3}}\left(|A\rangle_{sp} + |B\rangle_{sp} - |C\rangle_{sp}\right)$ as in the original 3-box paradox [6] (not to be confused with the "disappearing and reappearing" paradox described below).

According to Eq. 3 and the preparation above, before postselection the probe and shutter photon become entangled as follows:

$$|\Psi\rangle_{pp,sp} = \frac{1}{\sqrt{3}} \begin{bmatrix} \alpha_1 |RA\rangle_{pp}|A\rangle_{sp} + \alpha_2 |RB\rangle_{pp}|B\rangle_{sp} + \\ \alpha_1 |TA\rangle_{pp}\left(|B\rangle_{sp} + |C\rangle_{sp}\right) + \alpha_2 |TB\rangle_{pp}\left(|A\rangle_{sp} + |C\rangle_{sp}\right) \end{bmatrix}. \tag{4}$$

However, as shown in [6,7], the combination of pre- and post-selection implies that, if we look for the shutter photon in Box A/B, it will be found with certainty in A/B, respectively. While this holds for the time between pre- and post-selection, apparently making it inaccessible for verification, it becomes accessible by its entanglement with the probe photon (Eq. 4). Indeed, upon postselection of the shutter, the probe state while retaining its initial coherent superposition, becomes

$$|\text{Out}\rangle_{pp} = \alpha_1 |RA\rangle_{pp} + \alpha_2 |RB\rangle_{pp}, \tag{5}$$

which, in everyday language, reads: *a single probe photon is completely reflected by a single shutter photon retrodicted to have simultaneously resided in two boxes A and B.*

Here, quantum superposition (1) is revealed not only by the particle's passive passage along two paths, but by its active interaction with another particle in both locations (regarding the shutter's unique mode of presence in Box C, see discussion below). This interaction resembles Quantum Oblivion [9], where only one party seems to be affected by the interaction.

In what follows we discuss a fourth equation of this kind, revealing, in addition to the former three (1,2, and 5), a unique evolution in time. A particle, again operating as a shutter and superposed over three boxes, seems to abruptly disappear and reappear between them (see Fig. 1 and Fig. 2 for a detailed experimental scheme, which will be explained in detail in the following sections). This time-evolution is revealed by the probe photon being prepared with the normalized spatiotemporal superposition

$$|\text{In}\rangle_{pp} = \alpha_1 |A(t_1)\rangle_{pp} + \alpha_2 |C(t_1)\rangle_{pp} + \alpha_3 |C(t_2)\rangle_{pp} + \alpha_4 |B(t_3)\rangle_{pp} + \alpha_5 |C(t_3)\rangle_{pp}, \tag{6}$$

which means that it has a chance $|\alpha_1|^2$ to be in Box A at time $t_1$, $|\alpha_3|^2$ to be in Box C at time $t_1$ and so on. Thanks to their interaction (Eq. 3), we predict that this probe photon would be reflected by the pre- and post-selected shutter from *all* these varying positions in space and time. The probe photon can thus undergo interference of all its five spatiotemporal reflected trajectories

$$|\text{Out}\rangle_{pp} = \alpha_1 |RA(t_1)\rangle_{pp} + \alpha_2 |RC(t_1)\rangle_{pp} + \alpha_3 |RC(t_2)\rangle_{pp} + \alpha_4 |RB(t_3)\rangle_{pp} + \alpha_5 |RC(t_3)\rangle_{pp} \tag{7}$$

which, in everyday language, reads: *The shutter photon seems to reside in boxes A and C first, then only in C, and finally in B and C, reflecting the probe photon from all these varying locations.*

Here too, one would suggest alternative accounts for the position of the shutter photon based on local hidden variables instead of this extravagant disappearance-reappearance scenario. But here again, the above Bell-Hardy proof [2,4] for single-particle nonlocality rules out this alternative, thereby warranting an experimental realization for the explicitly nonlocal account straightforwardly given by Eq. 7. An even more straightforward proof is given in the following sections.

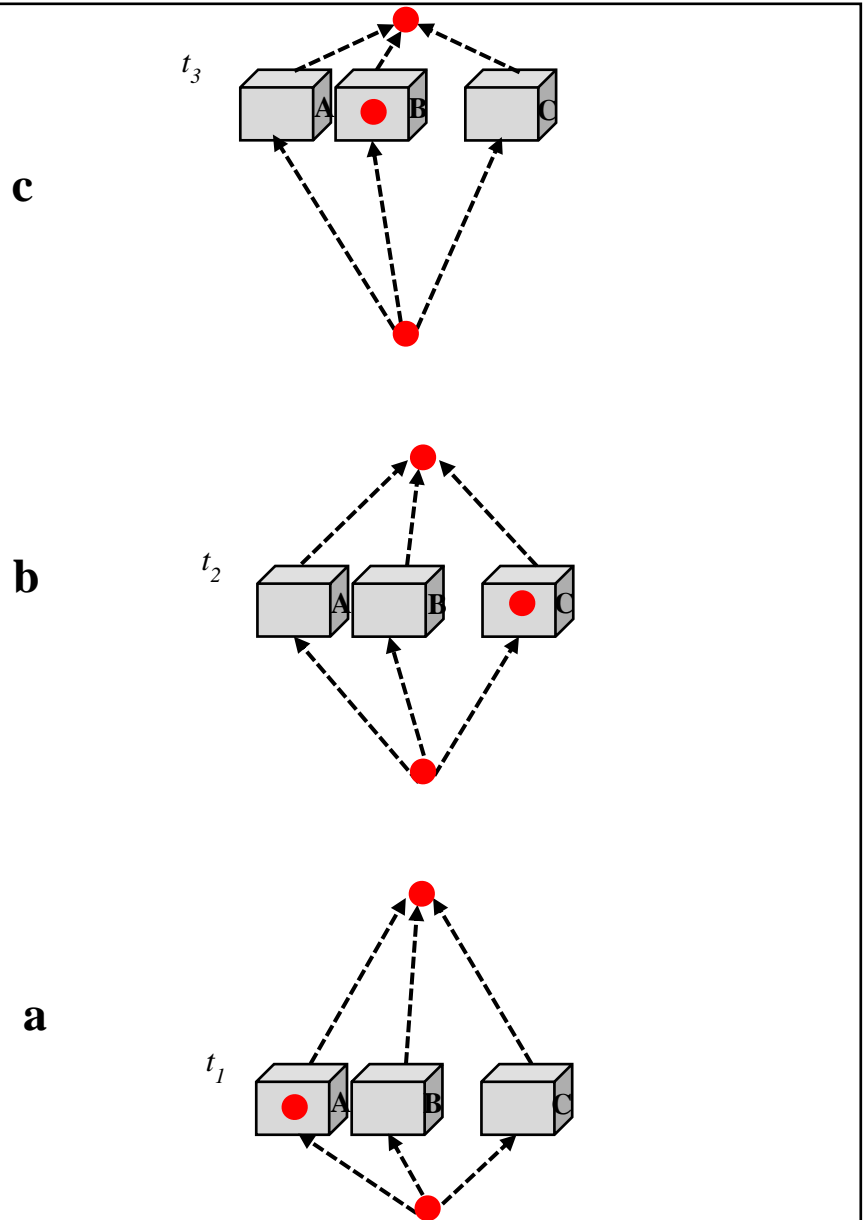

**Fig. 1**. **The time evolution of the disappearing particle experiment [22].** The combination of the two time evolutions gives rise to a curious disappearance and reappearance.

**The Two State-Vector Formalism.** For explaining the curious predictions of (5) and (7) we begin with a brief introduction to the specific formalism that has produced them. First, the Two-State-Vector Formalism (TSVF) [10-13] is fully consistent with the standard quantum formalism: All the former's predictions could equally be derived by the latter. Why, then, has that never happened? The main reason is the perfect time-symmetry underlying TSVF, so alien to quantum-mechanical intuition when measurement is involved. Once, however, TSVF *does* derive a prediction, it always turns out to be in full accord with quantum theory.

The TSVF predictions usually refer to the evolution of a particle between two consecutive quantum measurements. If, for instance, the particle is prepared at $t = t_i$ with $|\psi\rangle$ and later at $t = t_f$ found at a non-orthogonal state $|\phi\rangle$, then for all $t_i < t < t_f$ it can be described by the two-state $_{t_f}\langle\phi| \ |\psi\rangle_{t_i}$.

During this interval, and under special combinations of pre- and post-selected states, the particle is also expected to possess some intriguing properties, such as unusually small/large or even complex physical variables [14,15], or a spin separated from its mass [16], *etc.*

There is, however, an inherent obstacle for the validation of these predictions: The intermediate properties cannot, in general, be validated by an ordinary, projective measurement, because then it would *not* be a between-measurements state! This was the motivation for inventing weak measurements [17,18]: Let the coupling between particle and measuring apparatus be very weak, thereby highly inflicted by noise, but take many such measurements on an ensemble of particles (preferably in the same state) and average them out. Then, in compliance with basic statistical laws, the (constant) signal will strongly prevail over the (varying) noise, revealing the predicted phenomena without paying with the usual "collapse."

This technique, however, has invoked some skepticism, mainly related to its statistical character which allows more conservative explanations. This challenge merits consideration. If "The weight of evidence for an extraordinary claim must be proportioned to its strangeness" [19], it is *strong* projective measurement that should validate TSVF predictions.

**Advances: TSVF with Strong Measurements, Delayed Measurements, and the Bell-Hardy Nonlocality Proof.** These indeed were the advances made over the years, mostly recently. Experimental settings involving strong measurements were offered to test extraordinary TSVF predictions like the "Cheshire cat" [16], the "quantum pigeonhole" [20,21] and the "disappearing and re-appearing particle" [22,23]. Also, as noted above, a protocol involving strong measurements proposed by Aharonov and Vaidman [7] was experimentally realized by Okamoto and Takeuchi [8], to be elaborated in greater detail below.

One ingenious advance was made by an earlier work apparently not employing TSVF, yet doing just that. Hardy [4] has analyzed a single photon traversing an MZI and subtly interacting with two superposed atoms placed on its two paths. The photon is then selected for the cases where its interference is disturbed. Only then the two atoms, which interacted with the photon *earlier*, become EPR entangled. This setting belongs to the TSVF family for the simple reason that it addresses the photon's state which prevailed between two measurements, with the advance that the intermediate state is now subjected not to a weak but rather to a *delayed* strong measurement: The two atoms remain superposed and entangled till *after* the photon's post-selection, keeping the photon's earlier state "alive" even much later.

These advances are employed in what follows. Between the particle's pre- and post-selections, it is entangled with a photon, which later reveals the odd dynamics predicted for the intermediate time-interval.

**The Strong-Measurement Version of the Three Boxes Paradox and its Realization.** The three boxes paradox [6] is based on a counterfactual: "*Had* one measured the particle in only Box A, it would be there, and similarly for B" This is an extravagant type of nonlocality, where the very act of position measurement in a certain location seems to force the particle to "collapse" just there. Yet it is a *retrodiction*, holding only for a past state after the post-selection, which is by definition no longer accessible. This seems to rob the paradox much of its acuity.

This obstacle was overcome by Aharonov and Vaidman [7] in the general *N* boxes case, with a scheme analogous to Hardy's delayed measurement [4]. This time, then, it is a *strong* projective measurement of a particle operating as a "shutter." The experiment has two stages: *i*) The shutter particle, after the pre-selection, superposed over the three boxes, is coupled with a superposed probe particle. *ii*) Then, after the former's post-selection, the probe particle is subjected to a measurement which reveals the shutter particle's intermediate location. The retrodiction thus turns into a standard prediction: In all cases where the shutter's post-selection succeeds, the probe particle is reflected from all the *N-1* boxes, demonstrating the shutter's simultaneous existence in all of them.

Okamoto and Takeuchi [8] have recently tested this prediction of the TSVF for *N=3* in a quantum optics setup employing a novel photonic quantum router [24] which enables the shutter and the probe photons to interact. Their results show that the shutter did not randomly occupy one of the two boxes, *i.e.*, it did not "collapse." Rather (within experimental accuracy limitations), it has reflected the probe photon from both boxes. The latter was therefore measured (in a manner akin to interference) on both its possible return paths. Notice the novelty: Quantum superposition, so far demonstrated only by the particle's passive passage through multiple paths, is now made apparent with the particle's *active* operation on another particle, at all these locations.

**The Disappearing-Reappearing Particle.** We now come to our experiment. Whereas the three box paradox [6] presents an intriguing *state* – the particle reflecting the probe photon from two locations – the disappearing-reappearing particle Gedankenexperiment [22] gives an even more intriguing *evolution*: The particle is bound to be found in mutually-exclusive boxes at different times. And here again, the prediction is for strong (projective) measurements.

The pre-selection (in fact a preparation) is

$$|\psi(0)\rangle = \frac{1}{\sqrt{3}}(|A\rangle + i|B\rangle + |C\rangle), \qquad (8)$$

and the post-selection

$$|\phi(t_f)\rangle = \frac{1}{\sqrt{3}}(-|A\rangle + i|B\rangle + |C\rangle). \qquad (9)$$

We further introduce a time evolution $H = \varepsilon \sigma_x$ during $0 \leq t \leq t_f$, allowing the particle to move between Boxes A and B (according to this notation, $\sigma_x$ is the Pauli-x matrix, $|A\rangle \equiv |\uparrow\rangle$ and $|B\rangle \equiv |\downarrow\rangle$). Specifically, after a period of time denoted by $t_2$, a particle starting in A would be evenly superposed in A and B. After time $t_3$ this particle would completely move to Box B and after time $t_f$ it would return to A with a negative phase, but then we will project its state on Eq. 9. To simulate this evolution in our proposed photonic experiment, we would later use two beam-splitters.

Surprisingly, we have three different predictions (see also [22,23]) for the three instants during this interval, depending on the moment $t_1 \approx 0 / t_2 / t_3$ in which we decide to open the boxes and look for the particle:

$$\Pi_A(t_1) = 1 \text{ or } \Pi_C(t_1) = 1, \text{ conditioned on the choice which box to open,} \quad (10)$$

$$\Pi_A(t_2) = 0 \text{ and } \Pi_B(t_2) = 0 \text{ and } \Pi_C(t_2) = 1, \quad (11)$$

$$\Pi_B(t_3) = 1 \text{ or } \Pi_C(t_3) = 1, \text{ conditioned on the choice which box to open,} \quad (12)$$

where $\Pi_j \equiv |j\rangle\langle j|$ is the projection operator (in the Heisenberg picture) onto box $j = A, B, C$. These predictions are based on the corresponding weak values which coincide with the eigenvalues of the corresponding operators, and hence also imply the counterfactual results of a projective measurement, had it been performed [6].

In other words, a particle obliged by the formalism to reside in Box A at time $t_1$, is also obliged, with the same certainty, to "disappear" from it at time $t_2$, even though there is no tunneling from Boxes A and B to Box C. We note that if the values in Eqs. 9-11 are understood as weak values, they can be all weakly measured at the same experimental run [22]. Here, however, we present a *strong* simultaneous validation of these predictions, as follows.

It is time-symmetry that obliges this prediction. Followed from past to future, this formulation gives a reasonable account: Once you have prepared the pre-selected state (8), then, *if* you get either (10), (11) or (12), your probability to get the post-selection (9) goes up from 11% to 33%.

But the same evolution followed conversely (just as if pre- and post-selection were interchanged), is odd: If you did *not* perform any intermediate measurement, just obtained the pre- and post-selections (8) and (9), then *all three retrodictions (10,11,12) hold with 100% certainty*.

The TSVF explains this odd sequence of mutually-inconsistent locations as follows. In Eqs. 10-12 we focused on properties that can be found with certainty when performing a strong projective measurement. However, the weak values

$\langle \Pi_B(t_1) \rangle_w = \langle \Pi_A(t_3) \rangle_w = -1$, which can be validated through weak measurements, show that, in a sense, the total number of particles within Boxes A and B has been 0 all the time. Interestingly, at a time different from $t_2$ this 0 is achieved through a sum of two weak values – one positive and the other *negative*. Whereas a positive weak value indicates an ordinary effective interaction (although sometimes weaker or stronger than the one expected, due to weak values deviating from the spectrum of the measured operator), a negative weak value suggests an effective interaction with a minus sign. The particle therefore effectively behaves as if all its properties (such as mass and momentum) are negative, thereby termed *Nega-particle*. Notice that although the logic is that of weak values, in the present case they coincide with the eigenvalues of the projection operators, hence applicable also in the strong sense [6] which we examine here.

Moreover, at $t_2$ due to the tunneling between A and B described above, the particle's positive and negative weak values fully cancel each other, hence we expect its "disappearance." Then at $t_3$, when they part again, the particle "reappears." This happens continuously in time and in a self-consistent manner from the perspective of weak values.

**Bell's Inequality Ruling Out the "All Along" Alternative.** The above claims may seem unusual, even in comparison with other quantum effects. It would therefore be natural to consider a more moderate, even trivial alternative: The pre- and post-selections give mere *sub-ensemble*s. This would be what Griffiths denotes "a family of histories" [25], comprised of three *distinct* groups, each of which has its own properties:

i) Particles that went from the beam-splitter (BS) only to Box B at $t_1$ and then tunneled to A at $t_3$, therefore not yielding the post-selection (8). This family is ruled out by the post-selection.

ii) Particles that went to A at $t_1$ and then tunneled to B later, at $t_3$, yielding the post-selection (8) after being measured at either $t_1$ or $t_3$.

iii) Particles that went to C at $t_1$ and yield the post-selection (9) after being measured at any intermediate time.

Perhaps, then, no abrupt position changes occur because each individual particle has only one history. It may have travelled from A to B if found in either, but if found at any time in C, then it has been there *all along*. Similarly, if it is not found in A and B at $t_2$ it has been absent from them *all along*.

The flaw of this alternative is straightforward. Being "all along" somewhere is, by definition, a *local hidden variable*. Here the abovementioned Bell-Hardy proof [2,4] holds: Our single particle's wave-function, split into three, can exhibit nonlocal features, similar to those of an EPR pair or more specifically the W tripartite state [26] (although weaker in strength), this time apparently giving rise to a nonlocal disappearance and re-appearance. For, have we inserted three atoms into the three boxes in Eq. 8, as Hardy did with two atoms in [4], the single superposed photon would have entangled them in a W state (up to a relative phase which can be cancelled out) for which specific Bell-like inequalities are violated [27] implying the failure of local realism.

A similar shortcoming of the Consistent Histories formalism gave rise to the Entangled Histories formalism [28] and was recently described in [29,30].

Below, following the proposal for a laboratory realization of this experiment, we present a more straightforward Bell-type proof specifically adapted to the present shutter-probe photons pair.

**The Proposed Realization of the Disappearing-Reappearing Particle Experiment.**
Based on the Okamoto-Takeuchi [8] realization of Aharonov and Vaidman's [7] protocol, our proposal for a realization of the disappearing-reappearing particle naturally follows. It is based on the probe photon's being *superposed in time*, such as the time of emission/arrival, enabling temporal interference [31,32].

Preparing temporal superposition is easy. Split a single photon into three equal-intensity beams (see *e.g.* [33]) and delay the 2$^{nd}$ and 3$^{rd}$ beams. The three beams therefore pass through the 3-boxes system at $t_1$, $t_2$ and $t_3$, respectively.

At each instant, direct the corresponding 1/3 beam to one or more boxes in accordance with the TSVF prediction about the shutter-particle's position at that instant (Eqs. 10-12), as follows:

  i.    At $t_1$, the first 1/3 beam is split again, this time spatially, by a simple BS into 2, and goes to Boxes A+C.
  ii.   At $t_2$, the second 1/3 goes only to C.
  iii.  At $t_3$, the last 1/3, again split by simple BS, goes to B+C.

Then, after returning from all the boxes at all times, the probe is delayed and reunited in a manner precisely reverse to the above splits: first spatially and then temporally. Once all splits are completely undone, we measure the photon to see whether its initial quantum state is restored by interference.

Notice the required carefulness: To restore the photon's initial coherent superposition, a shutter must exist with certainty at all these boxes: (A+C)($t_1$), C($t_2$), (B+C)($t_3$). Failure of the shutter to be present at any box at the right instance would ruin the coherent superposition. Furthermore, there can be no other certain shutter positions during these times, *i.e.* this set of 5 slits closed by 1 shutter is maximal.

In addition to this direct test of the shutter-particle's presence, there is a complementary test, measuring its varying *absence*: Send a photon to A and B, the boxes predicted to be empty at $t_2$. This, conveniently, does not involve complicated momentum exchange between shutter and photon. Here too, only the specific combination of transmitted beams through (A+B)($t_2$), restores the photon's initial state with certainty.

**Experimental Setup and further Practical Considerations.** For the experimental implementation of this theoretical prediction, the realization of a quantum shutter is the most important task. For this purpose, we propose using a photonic quantum router (PQR) as shown in Fig. 2a [8]. The PQR consists of a two-mode nonlinear sign shift (NS) gate embedded in a Mach-Zehnder interferometer [8]. When there is no control photon input to mode $c_{in}$, the photon input to $a_{in}$ is routed to mode $a_{out}$. In contrast, the photon is routed to mode $b_{out}$ when a control photon is input to mode $c_{in}$.

Fig. 2b shows a photonic scheme using PQRs for our experiment. The input mode $b_{in}$ of PQR in Fig. 2a is omitted in Fig. 2b for visibility. In this scheme, the quantum shutter is represented by a shutter photon (SP), and the interaction between the quantum shutter and the probe photon (PP) is realized by PQRs. By passing through the beamsplitters and the phase shifter of $\pi/2$, the SP is prepared in the preselected state $\left(|A\rangle_{sp} + i|B\rangle_{sp} + |C\rangle_{sp}\right)/\sqrt{3}$ which corresponds to Eq. (8). Here, the continuous evolution in time as described by [4] is realized discretely; BS1 represents the tunneling between the boxes A and B between times $t_1$ and $t_2$. Similarly, BS2 is for the tunneling between times $t_2$ and $t_3$. Finally, SP is post-selected by the state $\left(-|A\rangle_{sp} - i|B\rangle_{sp} + |C\rangle_{sp}\right)/\sqrt{3}$ by passing through the beamsplitters and the phase shifters $(\pi, 3\pi/2)$. Note that the post-

selected state is slightly different from that in Eq. (9) because the time evolution from $t_3$ to $t_f$ can be embedded within the post-selected state, without affecting the result.

The PP is divided into 5 spatio-temporal beams, guided to five PQRs as follows. At $t_1$, while passing through PQRs, the first two PP beam interact with the SP in the modes (boxes) A and C. Similarly, PQRs provide the interactions between the PP and the SP in spatio-temporal modes C($t_2$), B($t_3$) and C($t_3$). As a result of the interaction at PQR, the PP is routed to the right output mode ($b_{out}$ in Fig. 2a) when the SP is in the same PQR. In contrast, when the SP is not input to PQR, the PP will be routed to the left mode ($a_{out}$ in Fig. 2a) and discarded (represented by "x" in the figures). After the interactions with the SP, the 5 spatio-temporal paths of the PP are re-merged by the beamsplitters. Since the pre- and post-selected SP appears in all the PQRs, the PP (due to interference) will be perfectly reflected with maintaining the coherence at each PQR [8]. As a result, the PP will be found at SPD2 with a probability of 100% when SP is detected at SPD1. Note that it is possible to reduce the number of the PQRs from 5 to 3 by recycling the PQR at mode C with an active optical path control.

To demonstrate disappearance and reappearance even simpler probe-shutter interactions suffice. Fig. 3 shows such a schemes for simplified, more feasible experiments. Fig. 3a shows a scheme for an experiment where the SP is in $|A\rangle_{sp}$ at time $t_1$, $|C\rangle_{sp}$ at time $t_2$ and $|B\rangle_{sp}$ at time $t_3$ can be confirmed. This scheme consists of three PQRs to check the reflections of PP from the space-time modes A($t_1$), C($t_2$) and B($t_3$). The total reflection brought about by the SP's disappearance and reappearance gives the unity detection probability of the PP at SPD2 when the SP is detected at SPD1. Fig. 3b further shows the simplest test for disappearance of the SP from box A, which has two PQRs to check the reflections from the two space-time modes A($t_1$), and C($t_2$). The unity detection probability at SPD2 proves the SP's absence from the mode A at $t_1$.

For NS gates, one can use heralding NS gates (single-mode-input) employing linear optics and single photons [34,35], which has been successfully demonstrated experimentally [36]. With the help of ancillary photons, the operation of these NS gates is 100% successful when a heralding signal is output. The two-mode-input NS gate required in Fig 2a is easily constructed using two single-mode-input NS gates [34-37] embedded in an interferometer. Therefore, in order to realize one PQR (Fig. 2a), two ancillary photons are required. Thus, the total number of 12 photons is required for implementing the scheme shown in Fig. 2b. The experiment using 12 photons seems technically difficult with current technology, but the recent rapid progresses on single photon sources [37,38] may make it possible in the near future. The feasibility of experimental implementation of simplified versions shown in Figs. 3a and 3b is much higher, because the total number of required photons for these schemes is 8 and 6, respectively.

Alternatively, one could also use post-selection-based NS gates [39-41], which are only successful when the number of incident photons from two input ports are kept at either of the two output ports. The advantage is that these gates do not require ancillary photons. However, one needs to know the number of photons at each of the outputs at every PQRs using nondestructive photon number measurement, which is technically very difficult. Otherwise, one could try to estimate the successful events using the photon detection at the final output ports and unconnected outputs of PQRs (shown by "x" in the figures) as done in a similar experimental demonstration [8].

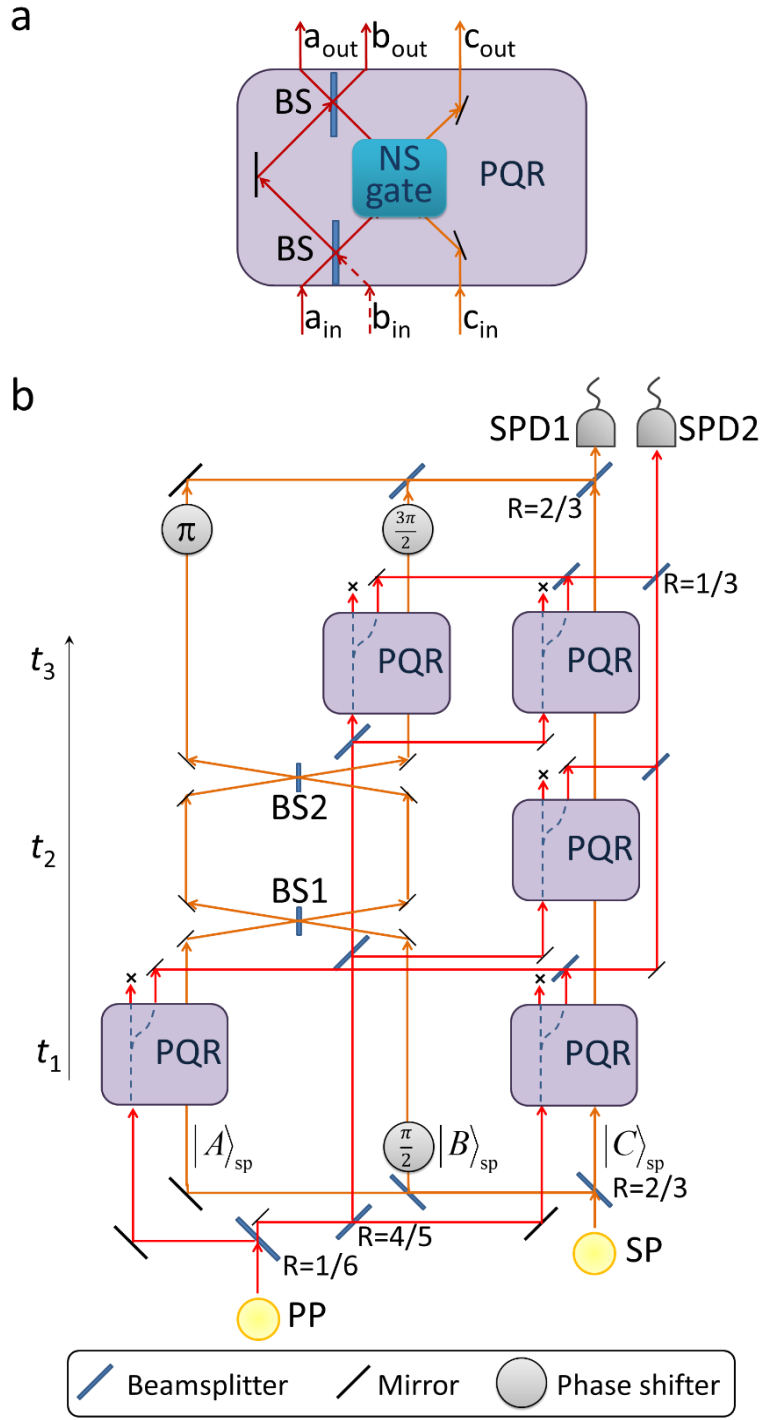

**Fig. 2. Photonic scheme using the photonic quantum routers for testing the predicted interactions between the Shutter (SP) and Probe (PP) photons (BS: beamsplitter; PQR: photonic quantum router; SPD: single photon detector).**
a. Photonic quantum router (PQR). A nonlinear sign shift (NS) gate is embedded in a Mach-Zehnder interferometer. B. The complete 5 spacetime-paths setting: PP is reflected by SP from boxes A+C at $t_1$, from C at $t_2$ and from B+C at $t_3$. The reflectance of the beamsplitters is 1/2 if not indicated.

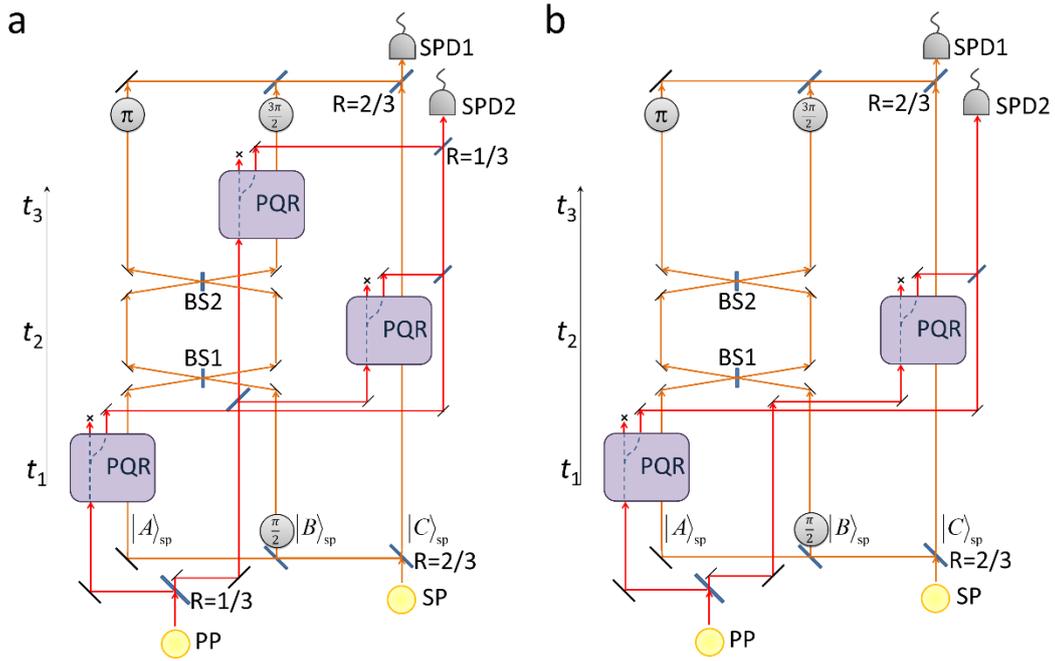

**Fig. 3. Simpler photonic schemes.**
a. A more feasible 3-paths setting with 3 channels: Reflections from A($t_1$), C($t_2$) and B($t_3$), showing the disappearance and re-appearance. b. The simplest test for disappearance: Reflections from A($t_1$), and C($t_2$). The reflectance of the beamsplitters is 1/2 if not indicated.

**A Stricter Version: Measuring Both Appearances and Disappearances with the Same Probe Photon.** The complex splitting of the probe photon in time and space, as well as the insertion of multiple PQRs, are feasible yet challenging, which is why a less complex laboratory version has been described above. Yet ignoring feasibility for a moment, an even more complex splitting is worth considering, where both the ideal proposal presented in Sec. "The Proposed Realization of the Disappearing-Reappearing Particle Experiment" and the control measurement at the end of that section are attained together.

To reveal both disappearance and reappearance it one setting, simply split the photon into six beams, such that even at $t_2$ it is spatially split into two. Now send these two rays at $t_2$ to Boxes A and B. The two PQRs in boxes A and B are now both switched in order to transmit the photon if no shutter photon is present. Here, the test is stricter: Any deviation from the sequence of transmissions, at $t_1$ and $t_3$ in the *presence* of a shutter and at $t_2$ in its *absence*, would fail to preserve the coherence of the probe photon (Fig. 4).

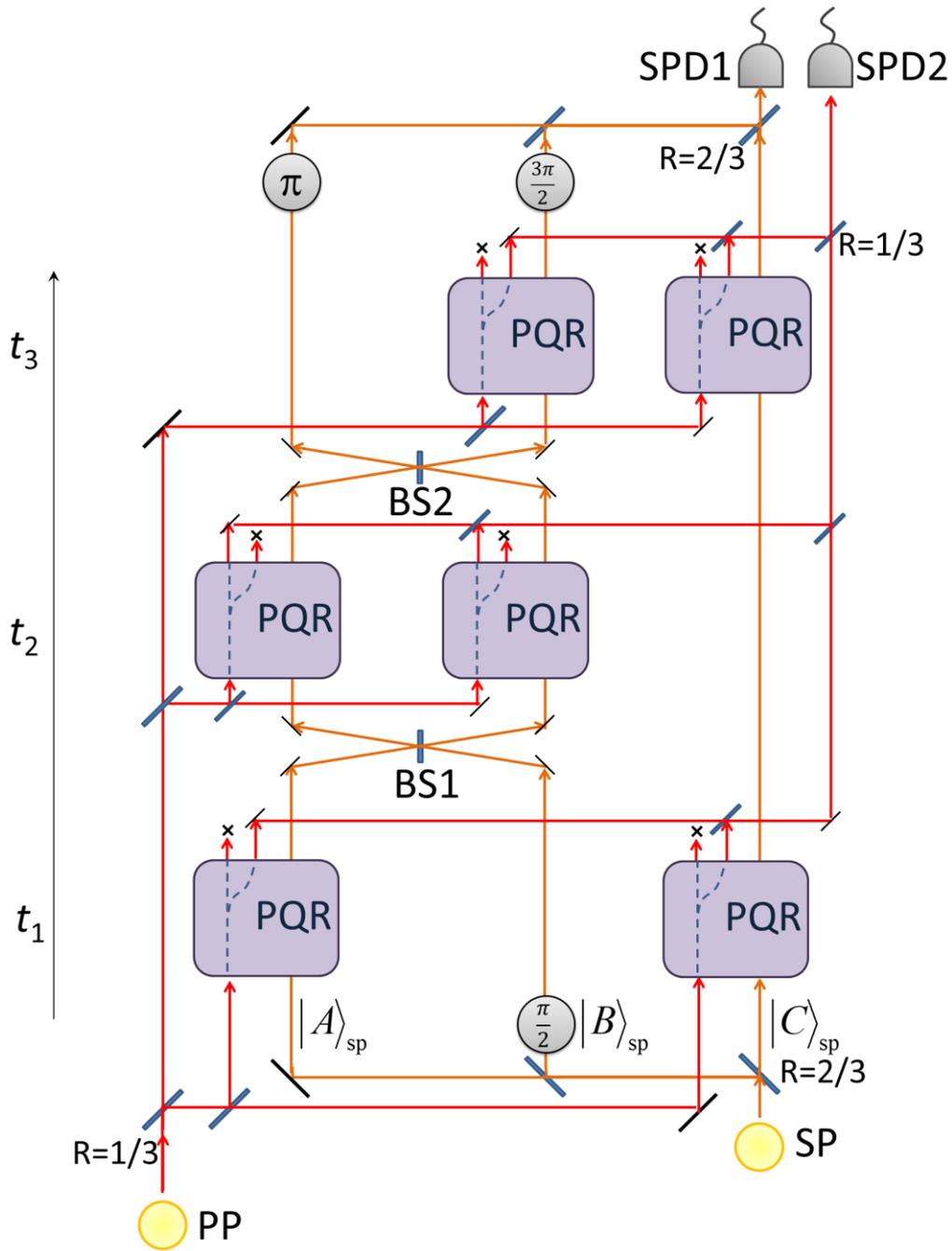

**Fig. 4. The particle's presence and absence measured by the same probe photon.** With a slightly more advanced setting, the probe photon, which is reflected by the shutter in the appropriate locations and times, measures also the shutter's absence in the intermediate time. At $t_2$, the PQRs in boxes A and B are switched so as to transmit this part of the probe photon's wave-function only if no shutter is present.

**An EPR-Bell-Type Validation.** Earlier, discussing the original disappearing-reappearing particle case with no measurement, we have employed the Bell-Hardy proof to rule out alternative interpretations to the predicted dynamics of the particle within the boxes. Now that two particles are involved, namely the shutter and probe photon, a more straightforward proof for the particles' nonlocal dynamics is at hand. It is based on the entangled state in Eq. 4 and its more complex spatio-temporal analogue preceding the outcome in Eq. 7. Okamoto and Takeuchi have already emphasized in [8] that the shutter and probe photons become, after interacting, an entangled pair. In our case, the state is not maximally entangled, containing a few terms that somewhat weaken the nonlocal correlations, but nevertheless we can employ the general theorem stating that "All entangled quantum states are nonlocal" [42]. This assures us that there does not exist any local hidden variables model which can account for the predicted outcomes.

On a more intuitive level, the probe-shutter entangled state prior to postselection is amenable to Bell-like tests. For this propose we may allow each particle to undergo either one of two different measurements, at the experimenters' choices. Following is a highly idealized scheme for this purpose: Let Alice and Bob collect the shutter and probe photons, respectively, prior to their unification. That is, Alice takes the three boxes A, B, and C where the shutter resides in superposition. Bob collects the reflected probe photon in five separate cavities, according to its reflection from the appropriate boxes at the appropriate times $i: A(t_1), ii: C(t_1), iii: C(t_2), iv: B(t_3), v: C(t_3)$. Each of them has the choice to either

- Open each of the boxes/cavities for a projective position measurement and find where the shutter/probe photon is;

or

- Reunite all of them and measure the shutter/probe photon for a non-commuting variable which is a coherent superposition of its possible positions (i.e. projecting on the complete superposed state with no relative phases). This measurement is akin to interference, measuring the photon's momentum.

The entangled state of the probe and shutter photons prior to postselection implies, according to [42], that each experimenter's choice of measurement nonlocally affects the outcome obtained by the other, in a way which cannot be reproduced using local hidden variables. But now, interestingly, the outcome is not a state but rather the particle's *entire evolution*.

Thus, our experiment entails (to various degrees depending on the $\alpha_j$ coefficients and the choices of measurements performed by the parties) nonlocal correlations between the probe and shutter photons. As such, the shutter's varying positions cannot be accounted for by any local hidden variables model.

**Discussion**

To summarize, we can better comprehend the bearing of the suggested experiment on the foundations of QM by considering again the simple state in Eq. 1, slightly adapted to the present context by splitting the wave-function into three:

$$|\psi\rangle = \frac{1}{\sqrt{3}}\left(|A\rangle + |B\rangle + |C\rangle\right). \tag{13}$$

Everyone has by now become accustomed to regarding this state as a fundamental quantum-mechanical state which allows a particle to equally reside in three places. Now, however, we have gained further insight into the nature of this superposition: Under a certain post-selection, the system's description is much more peculiar, namely the one pointed out by Aharonov *et al.* [6,7]. The particle would, in retrospect, appear in whatever location it is looked for – A or B. Apparently, the restriction "would, in retrospect" turns this prediction into a mere counterfactual, but this is not the case anymore. Thanks to Okamoto and Takeuchi's [8] experiment, we know that a delayed measurement can straightforwardly access this peculiar past state: Just let another probe particle, also split into two, interact with the shutter particle, but do not measure any particle yet. Then make the post-selection for the shutter. Finally, measure the probe-particle. The TSVF prediction, as experiment [8] has shown, can be now verified by the latter's coherence.

In this paper, then, an experiment is proposed to access an even more peculiar past state, in fact a past evolution: Under the appropriate pre- and post-selections, a delayed measurement should reveal the superposed particle's disappearance and re-appearance between distant boxes. Here too, a probe photon appropriately split to interact with the shutter at the appropriate times and locations can verify this unique dynamics.

Note, in passing, that this is also an example of the opportunities provided by pre- and post-selected quantum routers for photonic quantum computers. Further TSVF predictions with possible applications merit research along this line.

This work also sheds light on the controversial question regarding the past of a quantum particle [43,44]. It has been suggested in [43], and later tested in [44], that a particle has been located wherever it has left a weak trace, *i.e.* a non-zero weak value. In this paper we have demonstrated a stronger notion of a past – a particle has been located wherever it has acted as a shutter. As the terms suggest, a strong presence implies a weak one, but not vice versa. Unlike the former criterion suggested by Vaidman, this criterion can be tested with standard projective measurements and moreover, the single particle's past can, in theory, be verified with certainty.

Whenever the weak values do not coincide with the eigenvalues of a dichotomic operator, these predictions can be verified only with the aid of weak measurements. Nevertheless, this weak reality outlines a richer story underlying the predictions with strong (projective) measurements.

Let us then rephrase the question posed in the beginning: How literally should the TSVF retrodictions be taken? Our answer is based on two facts: *i*) TSVF is fully consistent with quantum theory. *ii*) Its counterfactual retrodictions have recently become ordinary predictions thanks to the introduction of delayed measurements. From these advances, a novel aspect of quantum uncertainty emerges. Perhaps "superposition" is actually a *collection* of many ontic states (or better, *two-time ontic states* [45]), which post-selection helps to isolate and measure under greater resolution. Some of these superpositions are even more unique, involving, *e.g.*, "collapse" of the particle's position into whatever box opened to find it [6], or possessing unusual momenta [46]. These phenomena can never be observed in real time, thereby avoiding violations of causality and other basic principles of physics. Yet the proof for their existence is as rigorous as the known proofs for quantum superposition and nonlocality – all of which are *post-hoc*. Very likely, other interesting forms of superposition are awaiting derivation and experiment, to advance our understanding of quantum reality.

**Acknowledgements**

It is a pleasure to thank Yakir Aharonov for many illuminating discussions. E.C. was supported by the Canada Research Chairs (CRC) Program. R.O. was supported by PRESTO, JST (No. JPMJPR15P4) and JSPS-KAKENHI (No. JP17H02936). S.T. was supported by JSPS-KAKENHI (No. JP26220712) and CREST, JST (No. JPMJCR1674).